\documentclass[12pt]{article}

\makeatletter \@addtoreset{equation}{section} \makeatother

\usepackage{epsfig,amsfonts}
\usepackage[fleqn]{amsmath}
\usepackage{amsthm,amssymb}
\usepackage{yfonts,mathrsfs}
\usepackage{graphicx}
\usepackage{hhline}
\usepackage{cite}
\usepackage{upgreek}
\def\be{\begin{equation}}
\def\ee{\end{equation}}
\def\bal{\begin{align}}
\def\eal{\end{align}}
\def\bea{\begin{eqnarray}}
\def\eea{\end{eqnarray}}

\def\nfrac#1#2{\genfrac{}{}{0pt}{}{#1}{#2}}

\def\pd{\partial}

\begin{document}


\begin{titlepage}
\begin{flushright}
RUNHETC-2008-24\\
\end{flushright}

\vspace{0.5cm}

\begin{center}
\begin{LARGE}
{\Large \textbf{On Correlation Numbers in $2D$ Minimal Gravity\\
and Matrix Models}}

\vspace{0.3cm}

\end{LARGE}

\vspace{0.5cm}

{\large Talk presented at "Liouville Gravity and Statistical
Models", International conference in memory of Alexei
Zamolodchikov, Moscow, June 21-24 2008}

\vspace{1.3cm}

\begin{large}

\textbf{A.A.Belavin$^1$, A.B. Zamolodchikov}$^{1,2}$

\end{large}

\vspace{1.cm}

${}^{1}$L.D. Landau Institute for Theoretical Physics\\
  Chernogolovka, 142432, Russia\\

\vspace{.2cm}

$\ {}^{2}$NHETC, Department of Physics and Astronomy\\
     Rutgers University\\
     Piscataway, NJ 08855-0849, USA\\

\vspace{1.0cm}

\centerline{\bf Abstract}

\vspace{.8cm}

\parbox{11cm}
{We test recent results for the four-point correlation numbers in
Minimal Liouville Gravity against calculations in the one-Matrix
Models, and find full agreement. In the process, we construct the
resonance transformation which relates coupling parameters
$\lambda_k$ of the Liouville Gravity with the couplings $t_k$ of
the Matrix Models, up to the terms of the order 4. We also
conjecture the general form of this transformation.}
\end{center}

\bigskip

\begin{flushleft}
\rule{4.1 in}{.007 in}\\
{November 2008}
\end{flushleft}
\vfill

\end{titlepage}
\newpage

\section{Introduction}

At present, there are two relatively independent approaches to
$2D$ Quantum Gravity. One is the continuous approach, in which the
theory is defined through the functional integral over the
Riemannian\footnote{Here we always have in mind the Euclidean
version of the theory.} metric $g_{\mu\nu}(X)$, with appropriate
gauge fixing. The choice of the conformal gauge leads to quantum
Liouville theory \cite{polyakov} (coupled to matter fields), and
for that reason this approach is often called the Liouville
Gravity. The other is the discrete approach, based on the idea of
approximating the fluctuating $2D$ geometry by an ensemble of
planar graphs, so that the continuous theory is recovered in the
scaling limit where the planar graphs of very large size dominate.
The discrete approach is usually referred to as the Matrix Models,
since technically the ensemble of the graphs is usually generated
by the perturbative expansion of the integral over $N\times N$
matrices, with $N$ sent to infinity to guarantee the planarity of
the graphs (see e.g. \cite{review} and references therein). Since
these two approaches stem from the same idea of fluctuating
geometry, they are expected to produce identical results for the
physical quantities. Indeed, this expectation was confirmed by
explicit calculations of some quantities in a number of particular
models \cite{kpz},\cite{gl},\cite{mpoint}. However, the techniques
involved in calculations by the two approaches are very different,
and satisfactory conceptual understanding of the relation between
those techniques is still lacking. In the absence of that, more
checks of the agreement of the results seem to be desirable.

The most easily-defined objects in the Quantum Gravity are the
$n$-point "correlation numbers" - the integrated correlation
functions
\begin{eqnarray}\label{cnumbers}
C_{k_1 ... k_n} = \langle \,O_{k_1} ... O_{k_n}\,\rangle\,, \qquad
O_k = \int_\mathbb{M}\,\mathcal{O}_k (X)
\end{eqnarray}
Here $\mathcal{O}(X)$ are some local densities (two forms) on the
manifold $\mathcal{M}$ which may involve both the "matter" and the
metric degrees of freedom localized at $X\in \mathbb{M}$, and the
expectation value is taken over the fluctuations of both the
matter and the geometry. The generating function
\begin{eqnarray}\label{zlambda}
Z(\{\lambda_k\}) = Z_0\,\langle\,\exp\big\{\sum_k
\,\lambda_k\,O_k\big\}\,\rangle
\end{eqnarray}
may be regarded as the vacuum partition function of the original
theory perturbed by adding the fields $\mathcal{O}_k (X)$ to the
action density, with the coupling constants $-\lambda_k$.
Throughout this paper we limit our attention to the case when
$\mathbb{M}$ is topologically a sphere. Even in this case, while
it is relatively easy to evaluate the $n$-point correlation
numbers in solvable Matrix Models, the Liouville Gravity approach
to these quantities involves technically complicated integration
over the $n-3$ dimensional moduli space of a sphere with $n$
punctures. For this reason, most previous comparisons of the
Matrix Models and the Liouville Gravity results was limited to
one-, two-, and three-point correlation numbers (Notable
exceptions are the analysis in \cite{mpoint}). A few years ago, a
new technique for handling the moduli integrals was developed by
Alexei Zamolodchikov and one of the authors \cite{bz}. It applies
to the so-called "Minimal Gravity" models (minimal CFT coupled to
the Liouville mode); by using the Higher Liouville Equations of
motion \cite{aliosha1} it allows one to reduce the moduli
integrals to the boundary terms. In this way the (partial) result
for the four-point correlation numbers in the Minimal Gravity was
obtained in \cite{bz}.

In this note we test the result of \cite{bz} against the
corresponding correlation numbers from the Matrix Models. There is
an important subtlety which makes this comparison less than
straightforward. In a local field theory, the integrated
correlation functions of the type \eqref{cnumbers} suffer from
intrinsic ambiguity generally referred to as the "contact terms".
Since \eqref{cnumbers} is the integral
\begin{eqnarray}\label{icorr}
\int_{X_1,...,X_n}\,\langle\,\mathcal{O}_{k_1}(X_1)...\mathcal{O}_{k_n}(X_n)\,
\rangle\,,
\end{eqnarray}
the correlation numbers may pick up contributions from delta-like
terms in the integrand, when two or more points $X_i$ collide.
Although most of the contact terms are not determined by the field
theory itself, it is well known that any change of contact terms
is equivalent to a certain analytic change of the coupling
parameters in \eqref{zlambda},
\begin{eqnarray}\label{ltrans}
\lambda_k\ \to\ {\tilde\lambda}_k = \lambda_k + \sum_{k_1 k_2}\,
C_{k}^{k_1 k_2}\,\lambda_{k_1}\lambda_{k_2} + \cdots
\end{eqnarray}
Of course, this relation is at the heart of the renormalization
theory, and indeed the ambiguity of the contact terms just
reflects the freedom of making finite renormalizations. Usually,
the ambiguity is resolved by appealing to the dimensional
analysis. The local fields $\mathcal{O}_k (X)$ can be chosen in
such a way that they (and hence the associated coupling parameters
in \eqref{zlambda}) have definite mass dimensions, so that
\begin{eqnarray}\label{deltak}
O_k \sim [\text{mass}^2]^{\delta_k}\,, \qquad \lambda_k \sim
[\text{mass}^2]^{-\delta_k}
\end{eqnarray}
Then, if one does not allow for any "auxiliary" dimensional
parameters, only transformations \eqref{ltrans} which respect the
balance of the dimensions are admitted, i.e. the non-linear (say,
$n$-th order) terms in \eqref{ltrans} are admissible only if they
satisfy the "resonance conditions"\footnote{The term is taken from
the vocabulary of the Renormalization Group, where precisely these
conditions characterize the nonlinear terms in the beta-functions
which can not be eliminated by analytic transformations of the
coupling parameters.}
\begin{eqnarray}\label{resonance}
\delta_k = \delta_{k_1} + ... + \delta_{k_n}
\end{eqnarray}
We generally refer to the non-linear terms in \eqref{ltrans} as
the "resonance terms".

In the absence of the resonances, no nonlinear finite
renormalizations \eqref{ltrans} consistent with the dimensional
counting are allowed, and thus the above ambiguity is completely
fixed. However, when the Minimal CFT is coupled to the gravity,
appearance of the resonances turns out to be very common.
Therefore this ambiguity needs to be resolved before a meaningful
comparison between the correlation numbers obtained by the two
different methods can be made. The problem was first outlined by
Moore, Seiberg, and Staudacher \cite{mss}, who also found the
solution in a number of special cases. In particular, addressing
the case of the $p$-critical point of the one-Matrix Model and the
corresponding Minimal Gravity $\mathcal{MG}_{2/2p+1}$ (the Minimal
CFT $\mathcal{M}_{2/2p+1}$ coupled to the Liouville Gravity), they
partly determined the resonance terms in the relation between the
coupling parameters, which allowed them to establish equivalence
up to the level of two-point correlation numbers.

In what follows we extend this analysis to three- and four-point
correlation numbers, and find perfect agreement between the Matrix
Models and Minimal Gravity results. In the process, we determine
the higher order resonance terms by demanding that the higher
order correlation numbers satisfy the fusion rules inherent to the
Minimal Gravity. At the end, we conjecture the full resonance
transformation which relates coupling parameters in the
$p$-critical one-Matrix Models and Minimal Gravity
$\mathcal{MG}_{2/2p+1}$.

\section{Minimal Gravity $\mathcal{MG}_{2/2p+1}$}

Specific models of the Liouville Gravity are defined by the
content of the "matter" field theory which is placed on the $2D$
manifold with the fluctuating metric. Perhaps the simplest models
are defined by choosing the Minimal CFT $\mathcal{M}_{q'/q}$ as
the "matter" theory. As in \cite{alz}, we use the term "Minimal
Gravity" $\mathcal{MG}_{q'/q}$ for such models. In this work we
restrict attention to the models with $q'=2$ and $q=2p+1$,
$p=1,2,3,...$. The Minimal Gravity $\mathcal{MG}_{2/2p+1}$ is very
likely to correspond to the $p$-critical point in the general
one-Matrix Model \cite{staudacher}.

\subsection{Minimal Model $\mathcal{M}_{2/2p+1}$}

The Kac table of the Minimal CFT $\mathcal{M}_{2/2p+1}$ is a
single raw of the length $2p$, the entries being the degenerate
primary fields $\Phi_{(1,n)}$, $n=1,2,...,2p$. Due to the
mandatory identification $\Phi_{(1,n)}=\Phi_{(1,2p+1-n)}$, the
model actually has only $p$ independent primary fields. We will
use the abbreviated notation
\begin{eqnarray}
\Phi_k = \Phi_{(1,k+1)}
\end{eqnarray}
for them, so that $\Phi_0$ is the identity operator. All the
independent primaries are listed by letting $k$ run through the
range $k=0,1,2,...,p-1$ (and in what follows, unless stated
otherwise, it is assumed that $k$ lies in this range), but it is
often convenient to extend the range to the full Kac table,
$k=1,2,...,2p$, with the above identification
\begin{eqnarray}\label{flip}
\qquad \Phi_{2p-k-1} = \Phi_{k}
\end{eqnarray}
kept in mind. In particular, with this convention the fusion rules
of $\mathcal{M}_{2/2p+1}$ take the simple form
\begin{eqnarray}\label{fusion}
[\Phi_{k_1}][\Phi_{k_2}] =
\sum_{k=|k_1-k_2|:\,2}^{k_1+k_2}\,[\Phi_k]
\end{eqnarray}
where as usual $[\Phi_k]$ stands for the irreducible Virasoro
representation associated with the primary field $\Phi_k$. Here
and below the symbol $\sum_{k=m:\,2}^{m'}$ denotes the sum with
the step 2, in which $k$ runs over the values $m,m+2,m+4,...,\leq
m'$. While in \eqref{fusion} it is assumed that $k_1, k_2$ lie
within the domain $[0,1,...,p-1]$, the summation index $k$ is
allowed to run outside it, where the terms $[\Phi_k]$ are
understood as $[\Phi_{2p-k-1}]$.

Note that the identification \eqref{flip} breaks the naive parity
symmetry $\Phi_k \to (-)^k\,\Phi_k$ of \eqref{fusion}, so that
$[\Phi_k]$ with odd (even) $k$ may appear in the r.h.s of
\eqref{fusion} with even (odd) $k_1+k_2$. In particular, the
correlation functions
\begin{eqnarray}\label{cfunctions}
\langle\,\Phi_{k_1}(X_1)\Phi_{k_2}(X_2) ...
\Phi_{k_n}(X_n)\,\rangle
\end{eqnarray}
do not necessarily vanish when $k_1+k_2+...+k_n$ is odd.

On the other hand, conformal invariance and the fusion rules
\eqref{fusion} do force many correlation function to vanish. Thus,
the conformal invariance demands vanishing of all the one-point
correlation function except for $\langle\,\Phi_0(X)\,\rangle$, as
well as the diagonal form of the two-point functions, i.e.
\begin{eqnarray}\label{onep}
&&\langle\,\Phi_{k}(X)\,\rangle =0 \qquad\qquad\qquad\,
\text{unless}
\quad k=0\,,\\
\label{twop}&&\langle\,\Phi_{k_1}(X_1)\Phi_{k_2}(X_2)\,\rangle =0
\qquad \text{unless}\quad k_1=k_2\,,
\end{eqnarray}
while the fusion rules \eqref{fusion} impose restrictions on the
multi-point correlation functions \eqref{cfunctions} with $n\geq
3$. Those restriction can be written in a compact form
\begin{eqnarray}\label{fusionn}
\eqref{cfunctions} = 0 \qquad \text{if} \quad
\bigg\{\nfrac{k_1+...+k_{n-1}\  <\ k_n \quad\quad\text{for}\quad
k_1+... +k_n \ \ \text{even}}{k_1+...+k_{n}\ < \ 2p-1
\quad\text{for}\quad k_1+... +k_n \ \ \text{odd}}
\end{eqnarray}
if one assumes that all $k_i$ are in the range $[0,1,...,p-1]$,
and that $k_n$ is the largest of them, i.e. $k_i \leq k_n$. Below
we generally refer to the case of even or odd $k_1+...+k_n$ as the
even and odd sectors.

\subsection{Coupling to Liouville Gravity}

According to \cite{polyakov}\cite{ddk}, coupling a conformally
invariant "matter" field theory to the fluctuating metric
$g_{\mu\nu}(x)$ leads, in the conformal gauge, to the famous
Liouville action
\begin{eqnarray}\label{laction}
\mathcal{A}_L = \frac{1}{4\pi}\,\int_\mathbb{M}\,\sqrt{\hat g}\
\left[{\hat g}^{\mu\nu}\,\pd_\mu\varphi\pd_{\nu}\varphi + Q\,{\hat
R}\,\varphi + 4\pi\mu\,e^{2b\,\varphi}\right]\,d^2 x
\end{eqnarray}
for the field $\varphi$ related to the conformal factor in
$g_{\mu\nu}=e^{2b\varphi}\,{\hat g}_{\mu\nu}$, where ${\hat
g}_{\mu\nu}$ is an arbitrary fixed "background" metric on
$\mathbb{M}$. The parameters $b$ and $Q$ are determined by the
central charge $c_M$ of the conformal "matter",
\begin{eqnarray}
26-c_M = 1+6\,Q^2\,\qquad Q=1/b+b\,.
\end{eqnarray}
In our case the "matter" is the Minimal CFT $\mathcal{M}_{2/2p+1}$
with $c_M = -6p+10-\frac{12}{2p+1}$, i.e.
\begin{eqnarray}
b=\sqrt{\frac{2}{2p+1}}\,.
\end{eqnarray}
The parameter $\mu$ in \eqref{laction} is dimensional,
\begin{eqnarray}
\mu \sim [\text{mass}]^2\,,
\end{eqnarray}
and interpreted as the cosmological constant.

The operators $O_k = \int_\mathbb{M}\,\mathcal{O}_k (x)$ are
constructed by "gravitational dressing" of the primary fields
$\Phi_k$,
\begin{eqnarray}\label{gdressing}
\mathcal{O}_k(x) = \Phi_k (x)\,\,e^{2a_k\,\varphi(x)}\,\sqrt{\hat
g}\ d^2 x\,,
\end{eqnarray}
where $a_k = b\,\frac{k+2}{2}$. This choice of the parameters
$a_k$ ensures that the integrand in \eqref{gdressing} is indeed a
density. These parameters also determine the mass dimensions of
the operators $O_k$, the $\delta_k$ in \eqref{deltak}
\cite{kpz}\cite{ddk},
\begin{eqnarray}\label{deltaka}
\delta_k = -\frac{a_k}{b} = -\frac{k+2}{2}
\end{eqnarray}

Proper gauge fixing leads to the following expression for the
correlation numbers \eqref{cnumbers} with $n\geq 3$
\begin{eqnarray}\label{npoint}
\langle O_{k_1}...O_{k_n}\rangle = \int_{x_1,\ldots,x_{n-3}}
\langle \mathcal{O}_{k_1}(x_1)...
\mathcal{O}_{k_{n-3}}(x_{n-3}){\tilde{\mathcal{O}}}_{k_{n-2}}(x_{n-2})
{\tilde{\mathcal{O}}}_{k_{n-1}}(x_{n-1})
{\tilde{\mathcal{O}}}_{k_{n}}(x_{n})
\rangle_{\nfrac{\text{Matter}\,\&}{\text{Liouville}}}\,,\quad
\end{eqnarray}
where $\mathcal{\tilde O}_{k} = C{\bar C}\,e^{2a_k
\varphi}\,\Phi_k$ are scalars (zero forms) associated with the
densities $\mathcal{O}_{k}$. Three points $x_{n-2},x_{n-1},x_{n}$
can be chosen at will, and the integration is performed over the
remaining $n-3$ points $x_1, \ldots , x_{n-3}$ (interpreted as the
moduli of the $n$-punctured sphere). The expectation value at the
right hand side of \eqref{npoint} involves both the "matter" and
the Liouville sectors. Due to the factorized form
\eqref{gdressing}, the integrand in \eqref{npoint} is a product of
the correlation functions \eqref{cfunctions} and the correlation
functions of the Liouville exponentials
\begin{eqnarray}\label{lexp}
\langle\,e^{2a_{k_1}\,\varphi(x_1)}\  \ldots \
e^{2a_{k_n}\,\varphi(x_n)}\, \rangle_{\text{Liouville}}
\end{eqnarray}

For $n=3$ no integration is necessary, and the three-point
correlation numbers are obtained by multiplying the three-point
functions \eqref{cfunctions} by the Liouville three-point function
\eqref{lexp}. The result takes a very simple form \cite{alz}
\begin{eqnarray}\label{3p}
\langle\,O_{k_1}O_{k_2}O_{k_3}\,\rangle =
-\mu^{\delta_{k_1}+\delta_{k_2}+\delta_{k_3}}\ N_{k_1 k_2 k_3}\
\mathcal{N}_p\ \prod_{i=1}^3\,\text{Leg}_L (k_i)\,.
\end{eqnarray}
Here
\begin{eqnarray}\label{np}
\mathcal{N}_p = (2p-1)(2p+1)(2p+3)\,,
\end{eqnarray}
and the "leg factors"
\begin{eqnarray}\label{legl}
{\rm Leg}_L (k) = \frac{(-)^\frac{k}{2}}{2\,\pi^\frac{k}{2}}\
\left[\gamma\left(\frac{2}{2p+1}\right)\right]^{-\frac{k+1}{2}}\,
\left[\gamma\left(\frac{2(k+1)}{2p+1}\right)\right]^\frac{1}{2}\,
\frac{\Gamma(p-1/2)}{\Gamma(p-k-1/2)}
\end{eqnarray}
($\gamma(t)=\Gamma(t)/\Gamma(1-t)$) will play a very small role in
our discussion\footnote{Note that $\text{Leg}_L (0) = 1/2$. It is
also interesting to note that for odd $k$ the factor \eqref{legl}
returns pure imaginary values. This reflects the known property of
the odd-$k$ operators $\Phi_k$ in the Minimal CFT
$\mathcal{M}_{2/2p+1}$. For example, $\mathcal{M}_{2/5}$ is known
to describe the Yang-Lee edge criticality \cite{cardy}; in this
identification the operator $\Phi_1$ corresponds to the Ising
order parameter, which couples to the pure imaginary magnetic
field.}. One can always get rid of them by multiplicative
renormalization of the operators $O_k$ and the coupling parameters
$\lambda_k$ in \eqref{zlambda}
\begin{eqnarray}\label{legout}
O_k \ \to\ \frac{1}{\text{Leg}_L (k)}\ O_k\,, \qquad \lambda_k \to
\text{Leg}_L (k)\ \lambda_k\,.
\end{eqnarray}
The most important entry in \eqref{3p} is the "fusion
coefficients" $N_{k_1 k_2 k_3}$ which take the value 1 if
three-point fusion rules are satisfied, and turns to zero
otherwise. Explicitly,
\begin{eqnarray}\label{f3odd}
N_{k_1 k_2 k_3} = \bigg\{\nfrac{1 \ \text{if}\ \ k_1+k_2+k_3 \geq
2p-1} {0\quad\qquad
\text{otherwise}\qquad\quad\quad}\qquad\text{for}\quad k_1+k_2+k_3
\ \ \text{odd}
\end{eqnarray}
and
\begin{eqnarray}\label{f3even}
N_{k_1 k_2 k_3} = \bigg\{\nfrac{1 \ \text{if}\ \ k_1+k_2 \geq k_3}
{0\qquad\ \  \text{otherwise}}\qquad\qquad\quad\ \ \ \
\text{for}\quad k_1+k_2+k_3 \ \ \text{even}
\end{eqnarray}
where in writing \eqref{f3even} we have assumed that $k_3$ is the
maximal of the numbers $k_i$, i.e. $k_1,k_2 \leq k_3$. This factor
of course is inherited from the "matter" three-point functions
$\langle\,\Phi_{k_1}\Phi_{k_2}\Phi_{k_3}\,\rangle$.

Since $\Phi_0$ is the identity operator in $\mathcal{M}_{2/2p+1}$,
insertion of $O_0$ in \eqref{cnumbers} is equivalent to taking the
derivative with respect to $\mu$, more precisely
\begin{eqnarray}
\langle\,O_0 \,O_{k_1}\ldots O_{k_n}\,\rangle =
-Z_{0}^{-1}\,\frac{\pd}{\pd
\mu}\,\left(Z_0\,\langle\,O_{k_1}\ldots O_{k_n}\,\rangle\right)\,,
\end{eqnarray}
 where
\begin{eqnarray}\label{zldim}
 Z_0 \sim \mu^\frac{2p+3}{2}
\end{eqnarray}
is the Liouville partition function of a sphere. Therefore the
two- and one-point correlation numbers are easily deduced from
\eqref{3p},
\begin{eqnarray}\label{p21}
\langle\,O_{k}O_{k'}\,\rangle = \delta_{k,\,\,k'}\
\frac{\mathcal{N}_p\ \mu^{2\delta_k}}{2p-2k-1}\ \text{Leg}_{L}^2
(k)\,, \qquad \langle\,O_k\,\rangle = -\delta_{k,\,0}\
(p+3/2)\,\mu^{-1}\,.
\end{eqnarray}
Again, the diagonal form of the two-point numbers and vanishing of
all but one of the one-point numbers can be traced to the
corresponding properties of the "matter" correlation functions,
Eq's \eqref{twop} and \eqref{onep}.

In Ref.\cite{bz} the integration over the moduli in \eqref{npoint}
is performed in the case $n=4$. The result, specialized for
$\mathcal{MG}_{2/2p+1}$, can be written as
\begin{eqnarray}
\langle\,O_{k_1}O_{k_2}O_{k_3}O_{k_4}\,\rangle = \mathcal{N}_p\
\mu^{\sum \delta_{k_i}}\  C_{k_1 k_2 k_3 k_4}\ \prod_{i=1}^4
\text{Leg}_{L}(k_i)
\end{eqnarray}
where, if one chooses $k_1$ to be the smallest of the numbers
$k_i$ (i.e. $k_1 \leq k_2,k_3,k_4$), the factor $C_{k_1 k_2 k_3
k_4}$ has the form
\begin{eqnarray}\label{p4}
C_{k_1 k_2 k_3 k_4}= (k_1+1)(p+k_1+3/2) - \sum_{i=2}^4\,
\sum_{s=-k_1:\,2}^{k_1}\,\bigg|p-k_i-s-\frac{1}{2}\bigg|\,.
\end{eqnarray}
An important qualification applies to this result. Eq.\eqref{p4}
was obtained under the assumption that the number of conformal
blocks in the decomposition of the $\mathcal{M}_{2/2p+1}$
correlation function
$\langle\,\Phi_{k_1}\Phi_{k_2}\Phi_{k_3}\Phi_{k_4}\,\rangle$ is
exactly $k_1$. To put it in explicit form, assume that $k_i$ are
arranged in non-increasing order,
\begin{eqnarray}\label{korder}
k_1 \leq k_2 \leq k_3 \leq k_4 \leq p-1\,.
\end{eqnarray}
Then, in the even sector, this assumption is fulfilled  iff
\begin{eqnarray}\label{14leq23}
k_1+k_4 \leq k_2+k_3
\end{eqnarray}
(which incidentally guarantees that the even-sector fusion rules
are satisfied). In the odd sector this condition also requires
that the fusion rules are satisfied with sufficient redundancy
\begin{eqnarray}
-k_1 +k_2 +k_3+k_4 \geq 2p-1\,,
\end{eqnarray}
which in turn demands validity of \eqref{14leq23} (since $k_4\leq
p-1$). Thus, in both even and odd sectors, when \eqref{14leq23}
breaks down, the validity of \eqref{p4} is questionable \cite{bz}.
Indeed, we will see in Section 4.2 that at $k_1 +k_4
> k_2+k_3$ \eqref{p4} deviates from the Matrix Model result.

As was mentioned in the Introduction, the Minimal Gravity
$\mathcal{MG}_{2/2p+1}$ is likely to be the world-sheet theory of
the $p$-critical point in the one-Matrix Model \cite{staudacher}.
This identification was confirmed by explicit comparison of the
one- and two-point correlation numbers \cite{mss}. The aim of this
paper is to extend the analysis to the higher order correlation
numbers, and in particular to test the new result \eqref{p4}
against the Matrix Models calculations.

\section{One-Matrix Model}

The matrix models technique was extensively studied in the
literature. Most of what we will need here can be found in
Ref.\cite{review} and references therein.

\subsection{$p$-critical point}

The one-Matrix Model exhibits an infinite set of multi-critical
points, labelled by the integer $p=1,2,3,...$. In the scaling
limit near the $p$-critical point, the partition function of the
sphere is expressed through the solution of the "string equation"
\begin{eqnarray}\label{se}
\mathcal{P}(u)=0\,,
\end{eqnarray}
where $\mathcal{P}(u)$ is the $p+1$-degree polynomial
\begin{eqnarray}\label{poly}
\mathcal{P}(u) = u^{p+1} + t_0\, u^{p-1} +\sum_{k=1}^{p-1}\,t_k\,u^{p-k-1}\,,
\end{eqnarray}
with the parameters $t_k$ describing the relevant deviations from
the $p$-critical point \footnote{Note that our labelling of the
parameters $t_k$ is different from that in Ref.\cite{mss}; our
$t_k$ are $t_{p-k-1}$ in \cite{mss}.}. The singular part of the
Matrix Model partition function $Z(t_0,t_1,...,t_{p-1})$ is
expressed through \eqref{poly} as follows\footnote{In the Matrix
Models technique it actually emerges through the equation
$\partial^2 Z/\partial t_{p-1}^2 = u_{*}(t_0,...,t_{p-1})$. The
Eq.\eqref{zmatrix} gives "physical" solution of this equation.},
\begin{eqnarray}\label{zmatrix}
Z = \frac{1}{2}\,\int_{0}^{u_{*}}\,\mathcal{P}^2 (u)\,du\,,
\end{eqnarray}
where $u_* = u_* (t_0,t_1,...,t_{p-1})$ is the suitably chosen
root of the polynomial \eqref{poly}, i.e. $\mathcal{P}(u_*)=0$. It
is important to remember that \eqref{zmatrix} really gives only
the singular part of the Matrix Model partition function. The
actual matrix integral has also a regular part, analytic in all
the parameters $t_k$ at $\{t_k\}=0$; in the spirit of the scaling
theory of criticality, the regular part is disregarded as
non-universal. For this reason the choice of $0$ as the lower
limit in the integral \eqref{zmatrix} is largely arbitrary. One
can replace it by any constant, or indeed any regular function of
$t_k$, at the price of adding the regular terms to $Z$.

There is strong evidence \cite{staudacher}, \cite{mss} that
\eqref{zmatrix} (with $u_*$ taken to be the maximal real root of
$\mathcal{P}(u)$) provides the Matrix Model description of the
Minimal Gravity $\mathcal{MG}_{2/2p+1}$, perturbed by the
operators \eqref{gdressing}. In this identification $-t_0$ is
interpreted as the cosmological constant $\mu$ in \eqref{laction}.
Obvious re-scaling symmetry $u \to a\,u,\  t_{k} \to
a^{k+2}\,t_{k}$ of Eq.\eqref{se} allows one to ascribe "mass
dimensions" to the the parameters $t_k$,
\begin{eqnarray}
t_k \ \sim \ [\text{mass}^2]^\frac{k+2}{2}
\end{eqnarray}
and then $Z \ \sim \ [\text{mass}^2]^\frac{2p+3}{2}$, in agreement
with \eqref{deltaka} and \eqref{zldim}.

Comparison at the level of the correlation numbers involves two
subtleties noticed long ago in Ref.\cite{mss}. First, it is
obvious from \eqref{deltaka} that at sufficiently large $p$ there
are many resonances \eqref{resonance} between the dimensions
$\delta_k$. Therefore the relation between the parameters $t_k$ in
\eqref{se} and the Minimal Gravity couplings $\lambda_k$ may
involve the resonance terms
\begin{eqnarray}\label{tlambda}
t_k = C_k\,\lambda_k + \sum_{n=2}^{\left[\frac{k+2}{2}\right]}\,
\sum_{k_1,..,k_n=0}^{p-1}\,C_{k}^{k_1 ... k_n}
\,\lambda_{k_1}...\lambda_{k_n}
\end{eqnarray}
with the coefficients constrained by the condition
\begin{eqnarray}
C_{k}^{k_1 ... k_n}=0 \quad\quad \text{unless} \quad\quad
\sum_{i=1}^n\,k_i = k+2-2n\,.
\end{eqnarray}
The sum over $k_i=0,1,...,p-1$ in \eqref{tlambda} includes $k_i=0$
to take into account the possibility that the integer powers of
the cosmological constant $\mu$ appear in the right-hand side; by
definition
\begin{eqnarray}
\lambda_0 = -\mu\,.
\end{eqnarray}
The coefficients $C_k$ have very little significance. They can be
removed by trivial renormalizations of the parameters $\lambda_k$
(or $t_k$) similar to \eqref{legout}. Physical equivalence between
the $p$-critical Matrix Model and the Minimal Gravity
$\mathcal{MG}_{2/2p+1}$ would imply that by special choice of the
coefficients in $C_{k}^{k_1 ... k_n}$ in \eqref{tlambda} the
partition function \eqref{zmatrix}, expressed through
$\{\lambda_k\}$, can be made identical to the generating function
\eqref{zlambda} of the Minimal Gravity, up to regular terms. This
was the idea put forward by Moore, Seiberg, and Staudacher
\cite{mss}, who have verified the identity up to the two-point
correlation numbers. In the next subsection we warm up by
re-deriving their result in somewhat different language, and then
proceed to the analysis of the higher orders.

The quantities to be compared with the correlation numbers of
$\mathcal{MG}_{2/2p+1}$ are the coefficients of the expansion
\begin{eqnarray}\label{zexp}
Z = Z_0 + \sum_{k=1}^{p-1}\,\lambda_k\ Z_k + \sum_{k_1
k_2=1}^{p-1}\, \frac{\lambda_{k_1}\lambda_{k_2}}{2}\ Z_{k_1 k_2} +
...+ \sum_{k_1,...k_n=1}^{p-1}\,\frac{\lambda_{k_1}...\lambda_{k_n}}
{n!}\ Z_{k_1 ...k_n} + ...\,.
\end{eqnarray}
of the Matrix Model partition function \eqref{zmatrix} in the
powers of $\lambda_1, ...\lambda_{p-1}$, with $\lambda_0=-\mu$
kept fixed. By dimensional analysis
\begin{eqnarray}\label{zkdim}
Z_{k_1 ... k_n} \ = \ z_{k_1 ... k_n}\
\mu^\frac{2p+3-2n-\sum\,k_i}{2}
\end{eqnarray}
with $z_{k_1 ... k_n}$ being just numbers. The coefficients with
even $\sum_{i=1}^n k_i$ (even sector) are half-integer powers of
$\mu$, and thus definitely belong to the singular part of the
partition function. However, $Z_{k_1 ... k_n}$ with odd
$\sum_{i=1}^n k_i$ (odd sector) involve integer powers of $\mu$.
When also
\begin{eqnarray}\label{oddreg}
\sum_{i=1}^n k_i \leq 2p+3-2n\,,
\end{eqnarray}
the odd-sector coefficients are non-negative powers of $\mu$, and
thus belong to the regular part of the partition function. As
such, they can be adjusted at will, and will be of no interest in
our analysis \footnote{However, note that these regular terms are
"special", in that they fully agree with the scaling (the
coefficients $z_{k_1 ... k_n}$ in \eqref{zkdim} are
dimensionless). Whereas generic regular terms of the full
microscopic partition function violate the scaling, and thus are
definitely beyond control of continuous field theory, there is
much reason to think that in Quantum Gravity the special regular
terms can be attributed some universal meaning (see
\cite{witten}). In this work we do not address this interesting
line of questions.}. Note that this inequality is always satisfied
for $n=1,2$, but at $n\geq 3$ negative powers of $\mu$ appear.
Therefore it is meaningful to compare the odd-sector correlation
numbers with $\sum k_i > 2p+3-2n$ with the results in
$\mathcal{MG}_{2/2p+1}$. We will return to this point in Sections
4 and 5 below.

\subsection{One- and two-point correlation numbers}

When one plugs \eqref{tlambda} into \eqref{poly}, the polynomial
takes the form
\begin{eqnarray}\label{polyk}
\mathcal{P}(u)=\mathcal{P}_0 (u) + \sum_{k=1}^{p-1}\,\lambda_k\,
\mathcal{P}_k (u) + ... +
\sum_{k_i=1}^{p-1}\,\frac{\lambda_{k_1}...\lambda_{k_n}}{n!}\,
\mathcal{P}_{k_1 ...k_n}(u) + ...
\end{eqnarray}
where $\mathcal{P}_0 (u)$ and $\mathcal{P}_{k_1 ... k_n}(u)$ are
the polynomials whose coefficients involve non-negative powers of
$\mu$. By dimensional analysis
\begin{eqnarray}\label{popk}
\nonumber&&\mathcal{P}_0 (u) = \ \ \ \ u^{p+1}\ \ \ +
C_{0}'\,\mu\,u^{p-1}
\ \ \ + C_{0}''\,\mu^2\,u^{p-3}\ \ \, + ...\\
&&\mathcal{P}_k (u) = C_k\,u^{p-k-1} + C_{k}'\,\mu\,u^{p-k-3} +
C_{k}''\,\mu^2\,u^{p-k-5} + ...\\
\nonumber &&...
\end{eqnarray}
(here $C_k$ are the same as in \eqref{tlambda}, and $C_{k}',
C_{k}'', ...$ are dimensionless constants related to the
higher-order coefficients in \eqref{tlambda}), and in general
$\mathcal{P}_{k_1 ... k_n}(u)$ are polynomials of the degree
\begin{eqnarray}\label{qorder}
p+1-2n-\sum k_i\,,
\end{eqnarray}
of similar structure. Of course, only polynomials of non-negative
degree appear, so that the sum in \eqref{polyk} is finite.

It is essential to notice that all the polynomials are either even
or odd,
\begin{eqnarray}\label{pksym}
\mathcal{P}_{k_1 ... k_n}(-u) = (-)^{p+1-\sum k_i}\
\mathcal{P}_{k_1 ... k_n}(u)\,,
\end{eqnarray}
because only integer powers of $\mu$ can appear in \eqref{popk}.
We use this symmetry to rewrite Eq.\eqref{zmatrix} in a somewhat
different form, more convenient for our analysis below. We split
the integration domain into two pieces
\begin{eqnarray}\label{intsplit}
\int_{0}^{u_*} = \int_{u_0}^{u_*} + \int_{0}^{u_0}\,,
\end{eqnarray}
where $u_0$ stands for the root of $\mathcal{P}_0(u)$ associated
with $u_*$, i.e. $u_*$ at $\lambda_1,...,\lambda_{p-1}=0$. Note
that
\begin{eqnarray}
u_0 = a_0\,\mu^\frac{1}{2}\,,
\end{eqnarray}
with some constant $a_0$ which plays no significant role in our
discussion below. The integrand in \eqref{zmatrix} involves the
products
\begin{eqnarray}\nonumber
\mathcal{P}_{k_1 ... k_m}(u)\mathcal{P}_{k_{m+1}...k_{n}}(u)
\end{eqnarray}
which are even or odd in $u$ depending on whether $k_1+...+k_n$ is
even or odd. For the even terms we can extend the integration in
the second piece in \eqref{intsplit} to the domain $[-u_0,u_0]$,
\begin{eqnarray}
\int_{0}^{u_0} \ \to \ \frac{1}{2}\,\int_{-u_0}^{u_0}
\end{eqnarray}
On the other hand, it is not difficult to see that the
contributions of the odd terms in that piece involve only
non-negative integer powers of $\mu$, and thus belong to the
regular part of the partition function. Thus, up to regular terms
\begin{eqnarray}\label{zmatrixm}
Z = \frac{1}{2}\,\int_{u_0}^{u_*}\,\mathcal{P}^2(u)\, du +
\frac{1}{4}\,\int_{-u_0}^{u_0}\,\mathcal{P}^2 (u)\,du\,,
\end{eqnarray}
and since the regular terms are of no interest to us here, in what
follows we study the partition function \eqref{zmatrixm}.

The first order of business is to determine $\mathcal{P}_0(u)$ and
$\mathcal{P}_k (u)$. Since $u_* = u_0 + O(\lambda_k)$, the first
term in \eqref{zmatrixm} does not contribute to the first three
orders in the expansion \eqref{zexp}, and one finds
\begin{eqnarray}
\label{z0}&&Z_0 =
\frac{1}{4}\,\int_{-u_0}^{u_0}\,\mathcal{P}_{0}^2 (u)\,du\,,\\
\label{zk}&&Z_{k}\ \ \ =
\frac{1}{2}\,\int_{-u_0}^{u_0}\,\mathcal{P}_0
(u)\mathcal{P}_k(u)\,du\,,\\
\label{zkk}&&Z_{k_1 k_2} = \frac{1}{2}\,
\int_{-u_0}^{u_0}\,\left[\mathcal{P}_{k_1} (u)\mathcal{P}_{k_2}(u)
+ \mathcal{P}_{0} (u)\mathcal{P}_{k_1 k_2}(u)\right]\,du\,.
\end{eqnarray}

Agreement with \eqref{p21} requires that all $Z_k$ vanish. Then
\eqref{zk} suggests that all the polynomials $\mathcal{P}_k (u)$
must be orthogonal to $\mathcal{P}_0(u)$ on the interval $[-u_0,
u_0]$, with the measure 1. Since the degrees of all
$\mathcal{P}_{k_1 k_2}(u)$ are smaller than $p-2$ (indeed, smaller
than $p-4$), the second term in \eqref{zkk} may be disregarded,
and then agreement with the diagonal form of the two-point
correlation numbers in \eqref{p21} requires that
$\mathcal{P}_{k}(u)$ themselves form an orthogonal set of
polynomials on that interval. One concludes that
$\mathcal{P}_k(u)$, up to normalization, are the Legendre
polynomials,
\begin{eqnarray}\label{pk}
\mathcal{P}_k(u)=C_k\,g_k\,u_{0}^{p-k-1}\ P_{p-k-1}(u/u_0)\,.
\end{eqnarray}
Here $C_k$ are the same as in \eqref{popk}, and
\begin{eqnarray}
g_k = \frac{(p-k-1)!}{(2p-2k-3)!!}\,.
\end{eqnarray}
We review some trivia about the Legendre polynomials in the
Appendix A (see e.g. \cite{be} for systematic display).
Furthermore, since $\mathcal{P}_0(u)$ is $p+1$ degree polynomial,
with no $u^p$ term, and vanishing at $u_0$, one finds
\begin{eqnarray}\label{p0}
\mathcal{P}_0 (u) =
g\,u_{0}^{p+1}\,\left[P_{p+1}(u/u_0)-P_{p-1}(u/u_0)\right]
\end{eqnarray}
where the normalization constant
\begin{eqnarray}
g = \frac{(p+1)!}{(2p+1)!!}
\end{eqnarray}
is inserted to guarantee that the highest order term in \eqref{p0}
is just $u^{p+1}$, as in \eqref{poly}. Incidentally, the
coefficient in front of $u^{p-1}$ in \eqref{p0} determines the
relation between $t_0$ in \eqref{poly} and the cosmological
constant,
\begin{eqnarray}
t_0 = -\frac{1}{2}\,\frac{p(p+1)}{2p-1}\,u_{0}^2
\end{eqnarray}

Eq.\eqref{z0} yields
\begin{eqnarray}
Z_0 = u_{0}^{2p+3}\ \frac{g^2\ (2p+1)}{(2p+3)(2p-1)}\,,
\end{eqnarray}
and then from \eqref{zkk}
\begin{eqnarray}\label{Zkk}
\frac{Z_{k k'}}{Z_0} = \delta_{k,\,\,k'}\ \frac{\mathcal{N}_p\
\mu^{-k-2}}{2p-2k-1}\ \text{Leg}_{M}^2 (k)\,,
\end{eqnarray}
where $\mathcal{N}_p$ is the same as in \eqref{np}, and
\begin{eqnarray}\label{legm}
\text{Leg}_{M}(k) = \frac{g_k\,C_k}{(2p+1)\,g\,a_{0}^{k+2}}\,.
\end{eqnarray}
Eq.\eqref{Zkk} reproduces the structure of the two-point numbers
in \eqref{p21}, and the identity of \eqref{legm} with the leg
factors \eqref{legl} fixes the normalization constants $C_k$.
Eq.\eqref{pk}, \eqref{p0} are equivalent to Eq. (4.24), (4.28) in
Ref.\cite{mss}\footnote{The relation
\begin{eqnarray}\nonumber
\int_{1}^\infty \,P_n (x)\,e^{-lx} = \sqrt{\frac{2}{\pi\,l}}\,\,
K_{n+1/2}(l)
\end{eqnarray}
between the Legendre polynomials and the Macdonald functions of
half-integer order makes the identity evident.}.

\section{Three- and four-point correlation numbers}

Before proceeding to the higher-order correlation numbers, it is
useful to get rid of annoying factors in \eqref{pk} and
\eqref{p0}. We trade $\lambda_k$ for the dimensionless couplings
\begin{eqnarray}\label{slambda}
s_k = \frac{g_k\,\,u_{0}^{-k-2}}{g\,(2p+1)}\,\lambda_k\,,
\end{eqnarray}
and write the polynomial \eqref{poly} as
\begin{eqnarray}
\mathcal{P}(u) = g\,(2p+1)\,u_{0}^{p+1}\,\,{Q}(u/u_0)\,,
\end{eqnarray}
where ${Q}(x)$ is the polynomial of degree $p+1$; as in
\eqref{polyk}, we will think of it as the power series in $s_k$,
\begin{eqnarray}\label{qexp}
{Q}(x) = {Q}_0 (x) + \sum_{k=1}^{p-1}\,s_k\,{Q}_k(x) + \sum_{k_1
k_2}^{p-1}\, \frac{s_{k_1}s_{k_2}}{2}\,{Q}_{k_1 k_2}(x) + ...
\end{eqnarray}
Eq's \eqref{p0} and \eqref{pk} then tell us that
\begin{eqnarray}\label{q0}
Q_0 (x) = \frac{P_{p+1}(x)-P_{p-1}(x)}{2p+1} = \int\,P_p (x)\,dx
\end{eqnarray}
and
\begin{eqnarray}\label{qk}
Q_k (x)=P_{p-k-1}(x)\,.
\end{eqnarray}

It is convenient also to trade the partition function
\eqref{zmatrix} for the dimensionless quantity $\mathcal{Z}
=\mathcal{Z}(s_1,...,s_{p-1})$
\begin{eqnarray}
Z = g^2\ (2p+1)^2\,u_{0}^{2p+3}\,\mathcal{Z}
\end{eqnarray}
given by
\begin{eqnarray}\label{zint}
\mathcal{Z} = \frac{1}{2}\,\int_{1}^{x_*}\,{Q}^2(x)\,dx +
\frac{1}{4}\,\int_{-1}^{1}\,{Q}^2(x)\,dx\,,
\end{eqnarray}
where $x_*=x_*(s_1,...,s_{p-1})$ is the largest real root of
${Q}(x)$. Note that $x_* (0,0,...,0)=1$, and
\begin{eqnarray}\label{qprim}
{Q}_{0}' (1) =1\,, \qquad {Q}_k (1)=1\,.
\end{eqnarray}

Up to the leg factors, the correlation numbers are the ratios
\begin{eqnarray}
\mu^{\sum \delta_{k_i}}\ \mathcal{Z}_{k_1 k_2 ...
k_n}/\mathcal{Z}_0
\end{eqnarray}
of the coefficients of the expansion
\begin{eqnarray}\label{rzexp}
\mathcal{Z} = \mathcal{Z}_0 + \sum_{k=1}^{p-1}\,s_k\ \mathcal{Z}_k
+ ...+ \sum_{k_1,...k_n=1}^{p-1}\,\frac{s_{k_1}...s_{k_n}} {n!}\
\mathcal{Z}_{k_1 ...k_n} + ...\,.
\end{eqnarray}
Note that with \eqref{q0}, Eq.\eqref{zint} yields
\begin{eqnarray}
\mathcal{Z}_0 = \frac{1}{4}\,\int_{-1}^{1}\,Q_{0}^2 (x)\,dx =
\mathcal{N}_{p}^{-1}\,,
\end{eqnarray}
where $\mathcal{N}_p$ is precisely the factor \eqref{np}.

\subsection{Three point numbers}

Evaluation of the coefficients $\mathcal{Z}_{k_1 k_2 k_3}$ is
straightforward; we put the calculations away into Appendix B. The
result is
\begin{eqnarray}\label{zkkk}
\mathcal{Z}_{k_1 k_2 k_3} = -1 +
\frac{1}{2}\,\int_{-1}^{1}\left[Q_{k_1 k_2}(x)Q_{k_3}(x) + Q_{k_1
k_3}(x)Q_{k_2}(x) + Q_{k_2 k_3}(x)Q_{k_1}(x)\right]\,dx\,,
\end{eqnarray}
The first term $-1$ reproduces \eqref{3p}, except for the fusion
rule factor $N_{k_1 k_2 k_3}$, Eq's \eqref{f3odd},\eqref{f3even}.
The role of the second term is to fix that discrepancy. Note that
in this case we need not worry about the odd sector. Recall from
\eqref{oddreg} that when $k_1 + k_2 + k_3$ is odd and $<2p-1$,
i.e. when the odd-sector fusion rules \eqref{f3odd} are violated,
the terms with $Z_{k_1 k_2 k_3}$ belong to the regular part of the
partition function. Therefore, we only need to look at the case
when $k_1 +k_2 +k_3$ is even, where \eqref{f3even} demands that
the second term in \eqref{zkkk} turns to $1$ at all configurations
of $k_1, k_2, k_3$ such that $k_1+k_2 > k_3$ (as in \eqref{f3even}
we assume that $k_1,k_2 \leq k_3$), to cancel the first term in
\eqref{zkkk}. To reproduce the fusion rule factor $N_{k_1 k_2
k_3}$ we need to have
\begin{eqnarray}
\frac{1}{2}\,\int_{-1}^{1}\,Q_{k_3}(x)Q_{k_1 k_2}(x)\,dx =\bigg\{
\nfrac{1\quad\text{if}\quad k_1+k_2<k_3}{0\quad\text{if}\quad
k_1+k_2 \geq k_3}
\end{eqnarray}
Since $Q_k (x) = P_{p-k-1}(x)$, this is achieved by taking (see
Eq.\eqref{Ap1p} in the Appendix A)
\begin{eqnarray}\label{qkk}
Q_{k_1 k_2}(x) = P_{p-k_1-k_2-2}'(x)\,,
\end{eqnarray}
where prime denotes the derivative of the Legendre polynomial with
respect to $x$. Note that now for some admissible values of
$k_1,k_2$ the index in \eqref{qkk} can take negative values;
throughout this paper we adopt the convention that $P_n(x)$ with
negative $n$ are identically zero. With \eqref{qkk}, we have
\begin{eqnarray}
\mathcal{Z}_{k_1 k_2 k_3}/\mathcal{Z}_{0}= - N_{k_1 k_2
k_3}\,\mathcal{N}_p\,,
\end{eqnarray}
in exact agreement with \eqref{3p}.

\subsection{Four point numbers}

Direct calculation (Appendix B) yields
\begin{eqnarray}\label{zkkkk}
\mathcal{Z}_{k_1 k_2 k_3 k_4} = \mathcal{Z}_{k_1 k_2 k_3
k_4}^{(0)} + \mathcal{Z}_{k_1 k_2 k_3 k_4}^{(\text{I})}\,,
\end{eqnarray}
where
\begin{eqnarray}\label{zkkkk0}
\mathcal{Z}_{k_1 k_2 k_3 k_4}^{(0)} = - F(-2) +\sum_{i=1}^4
\,F(k_i-1) - F(k_{(12|34)}) - F(k_{(13|24)}) - F(k_{(14|23)})\,,
\end{eqnarray}
and
\begin{eqnarray}\label{zkkkkI}
\mathcal{Z}_{k_1 k_2 k_3 k_4}^{(\text{I})} =
\frac{1}{2}\,\int_{-1}^{1}\,\left[Q_{k_1 k_2 k_3}Q_{k_4} + Q_{k_4
k_1 k_2}Q_{k_3}+Q_{k_3 k_4 k_1}Q_{k_2}+Q_{k_2 k_3
k_4}Q_{k_1}\right]\,dx \,.&&
\end{eqnarray}
In \eqref{zkkkk0}
\begin{eqnarray}
F(k)=P_{p-k-2}' (1) =
\frac{1}{2}\,(p-k-1)(p-k-2)\,\Theta_{p-2,k}\,,
\end{eqnarray}
with $\Theta_{k,k'}$ being the step function
\begin{eqnarray}\label{step}
\Theta_{k,k'} = \bigg\{\nfrac{1 \quad \text{for} \quad k\geq k'}
{0 \quad \text{for} \quad k<k'}
\end{eqnarray}
and we use the notation
\begin{eqnarray}
k_{(ij|lm)} = \min(k_i+k_j,k_l+k_m)\,.
\end{eqnarray}
Like in \eqref{zkkk}, the role of the term \eqref{zkkkkI} is to
enforce the fusion rules, and the polynomials $Q_{k_1 k_2 k_3}(x)$
are to be determined from this requirement.

The analysis is more simple in the even sector, so let's start
with this case. Assume again that the numbers $k_1, k_2, k_3, k_4$
are arranged as in \eqref{korder}, so that in \eqref{zkkkk0} we
always have
\begin{eqnarray}\label{kk1213}
k_{(12|34)} = k_1+k_2\,, \qquad k_{(13|24)} = k_1 +k_3\,.
\end{eqnarray}
Recalling \eqref{qk}, and counting the degrees of the polynomials
$Q_{k_1 k_2 k_3}(x)$, one observes that \eqref{zkkkkI} vanishes
when the even sector fusion rules \eqref{fusionn} are satisfied.
On the other hand, when the fusion rules are violated, the
inequality \eqref{14leq23} is violated as well, and we have
\begin{eqnarray}\label{kk23}
k_{(14|23)}= k_2+k_3 < p-1
\end{eqnarray}
where the last inequality follows from $k_4 \leq p-1$. With
\eqref{kk1213} and \eqref{kk23} the expression \eqref{zkkkk0}
evaluates to
\begin{eqnarray}\label{zkkkk0b}
-\frac{1}{2}\,(k_4-k_1-k_2-k_3-2)(2p-3-k_1-k_2-k_3-k_4)\,.
\end{eqnarray}
Thus, for \eqref{zkkkk} to satisfy the even-sector fusion rules
\eqref{fusionn} the integral
\begin{eqnarray}
\int_{-1}^{1}\,Q_{k_1 k_2 k_3}(x) Q_{k_4}(x)\,dx
\end{eqnarray}
has to return \eqref{zkkkk0b} with the opposite sign. This
uniquely determines the polynomials $Q_{k_1 k_2 k_3}$,
\begin{eqnarray}\label{qkkk}
Q_{k_1 k_2 k_3}(x) = P_{p-\sum k_i -3}''(x)
\end{eqnarray}
(see Eq.\eqref{Ap2p} in Appendix A).

Now, in the odd sector, as was explained in Sect.2.2, the
coefficients \eqref{zkkkk} with $\sum k_i \leq 2p-5$ correspond to
regular terms in the partition function $Z$, and thus can be
disregarded. For that reason, in particular, we can pay no
attention to the second term, Eq.\eqref{zkkkkI}. The only singular
terms potentially violating the odd-sector fusion rules
\eqref{fusionn} are those with
\begin{eqnarray}\label{evensing}
\sum k_i = 2p-3\,.
\end{eqnarray}
But it is easy to check that these terms actually vanish. Indeed,
if again $k_i$ are arranged as in \eqref{korder}, and $k_1+k_4
\leq k_2+k_3$ (in which case it follows from \eqref{evensing} that
also $k_1+k_4 \leq p-1$), the \eqref{zkkkk0} is given by
\eqref{zkkkk0b} which vanishes at \eqref{evensing}. If instead
$k_2+k_3\leq k_1+k_4$ (and hence $k_2+k_3\leq p-1$), the
expression \eqref{zkkkk0} reduces to
\begin{eqnarray}\label{zkkkk0a}
(k_1+1)(2p-3-k_1 -k_2 -k_3 -k_4)\,,
\end{eqnarray}
again vanishing at \eqref{evensing}.

Once the fusion rules are taken care of, we can compare
\eqref{zkkkk} with the $\mathcal{MG}_{2/2p+1}$ four-point
correlation number. It is not difficult to check that when
\eqref{14leq23} is satisfied, \eqref{zkkkk0} reproduces \eqref{p4}
perfectly.

However, when instead $k_2+k_3 < k_1 +k_4$, \eqref{zkkkk0} no
longer agrees with \eqref{p4}. This is expected, since the
Liouville Gravity calculations in \cite{bz} were made under a
certain assumption which is violated in this case. The Matrix
Model result \eqref{zkkkk}, which is valid at all configurations
of $k_i$, may provide a hint on how to modify the Liouville
Gravity analysis when the number of conformal blocks in the
$\mathcal{M}_{2/2p+1}$ four-point function is smaller then $k_1$.

\section{Multi-Point correlation numbers}

In principle, one can extend the analysis of the previous Section
to $n$-point correlation numbers with $n>4$. Although at the
moment no specific results for these multi-point numbers in the
Minimal Gravity are available, one thing is known. Since the
correlation functions \eqref{cfunctions} of $\mathcal{M}_{2/2p+1}$
vanish when the fusion rules \eqref{fusionn} are violated, the
correlation numbers \eqref{npoint} then vanish as well. This
requirement for the $n$-point numbers imposes strong conditions on
the form of the polynomials $Q_{k_1 ... k_{n-1}}(x)$, which fix
them uniquely, and thus one must be able to determine the complete
form of the polynomial $Q(x)$, Eq.\eqref{qexp}, step by step in
$n$.

In fact, the problem seems over-determined. Indeed, suppose we
have already constructed the expansion \eqref{qexp} up to the
order $n-1$, and thus $Q_0, Q_k, ..., Q_{k_1 ...k_{n-2}}$ are
already determined. Then $Q_{k_1 ...k_{n-1}}$ enters the
expression for the $n$-th order coefficient $\mathcal{Z}_{k_1
...k_{n}}$ only through the "counterterm"
\begin{eqnarray}\label{ncounter}
\frac{1}{2}\,\int_{-1}^{1}\,Q_{k_1 ... k_{n-1}}(x)Q_{k_{n}}(x)\,dx
\end{eqnarray}
and $n$ similar terms which differ from this by cyclic
permutations of $k_1,...,k_{n}$. The polynomials $Q_{k_1 ...
k_{n-1}}$ must be chosen in such a way that these terms cancel all
other contributions to $\mathcal{Z}_{k_1 ...k_{n}}$ when the
even-sector fusion rules \eqref{fusionn} are violated, i.e. when
$k_1+...+k_{n-1}>k_n$. But since the degree of the polynomial
$Q_{k_1 ...k_{n-1}}(x)$ is $p+3-2n-(k_1+...+k_{n-1})$ (see
Eq.\eqref{qorder}), the integral \eqref{ncounter} actually
vanishes at $k_1+...+k_{n-1} > k_n+4-2n$. For $n\geq 4$ a window
\begin{eqnarray}\label{weven}
k_{n}\  >\  \sum_{i=1}^{n-1}\,k_i \ >\  k_n +4 -2n
\end{eqnarray}
opens in configurations of $k_i$ violating the even-sector fusion
rules, where the term \eqref{ncounter} seems to be incapable of
doing its job of fixing the fusion rules. A similar problem exists
in the odd sector. For $n\geq 4$ there is a window
\begin{eqnarray}\label{wodd}
2p-1\ >\ \sum_{i=1}^{n}\,k_i\ > \ 2p + 3 -2n
\end{eqnarray}
in the configurations of $k_i$, where the odd-sector fusion rules
\eqref{fusionn} are violated, but corresponding coefficients
$Z_{k_1 ... k_n}$ are singular (involve negative integer powers of
$\mu$, see Sect. 3.1).

We have seen in Section 4.2 that at $n=4$ the problem takes care
of itself, in both even and odd sectors. The first factor in
Eq.\eqref{zkkkk0b} forces the coefficients $Z_{k_1 k_2 k_3 k_4}$
to vanish within the window \eqref{weven}, and the last factor in
\eqref{zkkkk0b} and \eqref{zkkkk0a} makes sure that they vanish
within the window \eqref{wodd} as well.

The same phenomenon should persist for higher $n>4$, for otherwise
we would face incurable disagreement between the correlation
numbers in $\mathcal{MG}_{2/2p+1}$ and the Matrix Model. We have
calculated the five-point correlation numbers $C_{k_1 k_2 k_3 k_4
k_5}$, and indeed they automatically vanish within both even and
odd sector windows, Eq.\eqref{weven} and Eq.\eqref{wodd}. We do
not present this calculation here, for the result is somewhat
cumbersome, and anyway at the moment there is nothing on the
Minimal Gravity side to compare it with. But as the byproduct of
this calculation we have determined the four-index polynomials
$Q_{k_1 ... k_4}$,
\begin{eqnarray}\label{qkkkk}
Q_{k_1 k_2 k_3 k_4}(x) = P_{p-\sum k -4}'''(x)\,,
\end{eqnarray}
where $\sum k = k_1 + k_2+k_3+k_4$.

\section{Discussion}

Identification of $\mathcal{MG}_{2/2p+1}$ as the world-sheet
theory of the $p$-critical one-Matrix Model suggests that, by
choosing suitable resonance terms in the of the relation between
the couplings $t_k$ in \eqref{poly} and $\lambda_k$, the Matrix
Model correlation numbers can be put in agreement with the fusion
rules of $\mathcal{MG}_{2/2p+1}$. Technically, this is done by
constructing the polynomial $Q(x)$, Eq.\eqref{qexp}, order by
order in $s_k$. In Sections 4 and 5 above, we have executed this
program up to the fourth order.

For higher $n$ direct calculations become rather involved. But a
quick glance at \eqref{qk},\eqref{qkk}, \eqref{qkkk}, and
\eqref{qkkkk} immediately suggests the general form,
\begin{eqnarray}
Q_{k_1 ... k_n}(x) = \left(\frac{d}{dx}\right)^{n-1} P_{p-\sum k
-n}(x)\,,
\end{eqnarray}
where again $\sum k = k_1 + ... +k_n$. Then, using \eqref{Apint},
the full polynomial $Q(x)$ can be neatly written as the integral
\begin{eqnarray}\label{intq}
Q(x) = -
\oint_{0}\,\frac{\left(1-2xz+z^2-\sum_{k=1}^{p-1}\,2s_k\,z^{k+2}
\right)^{1/2}}{z^{p+2}}\,\frac{dz}{2\pi i}\,,
\end{eqnarray}
where the integration is over a small circle around $0$. So far,
we did not find a proof that all correlation numbers computed from
\eqref{zint} with $Q(x)$ given by \eqref{intq} obey all fusion
rules \eqref{fusionn}. Thus, \eqref{intq} remains a conjecture.

\section*{Appendix}

\appendix

\section{Legendre Polynomials}

The Legendre polynomials $P_n(x)$ are $n$-th order polynomials
which form an orthogonal system on the interval $[-1,1]$ with the
weight 1,
\begin{eqnarray}\label{lr0}
\int_{-1}^{1}\,P_{n}(x)P_{n'}(x)\,dx =
\frac{2\,\delta_{n,n'}}{2n+1}\,.
\end{eqnarray}
The standard normalization is such that
\begin{eqnarray}
P_n (1) =1\,.
\end{eqnarray}
Explicitly,
\begin{eqnarray}
P_n (x) = \frac{2^{-n}}{n!}\ \frac{d^n}{dx^n}\left[x^2-1\right]^n
= 2^{-n}\,\sum_{l=0}^{[n/2]}\, (-)^l\,\frac{(2n-2l)!}{l!
(n-l)!(n-2l)!}\,\,x^{n-2l}\,.
\end{eqnarray}
Another closed expression is given in terms of the Hypergeometric
series,
\begin{eqnarray}
P_n (x) =\  _2 F_1 \left(-n,n+1,1;\frac{1-x}{2}\right)\,,
\end{eqnarray}
i.e.
\begin{eqnarray}
P_{n}'(1)=\frac{n(n+1)}{2}\,, \quad
P_{n}''(1)=\frac{(n-1)\,n\,(n+1)(n+2)}{8}\,, \quad \text{etc}
\end{eqnarray}
Yet another closed form is in terms of the contour integral
\begin{eqnarray}\label{Apint}
P_n(x) =
\oint_{0}\,\frac{\left(1-2xz+z^2\right)^{-1/2}}{z^{n+1}}\,
\frac{dz}{2\pi i}\ .
\end{eqnarray}

The following relations are useful in our analysis:
\begin{eqnarray}\label{lr1}
P_{n+1}'(x)-P_{n-1}'(x) = (2n+1)\,P_n(x)\,;
\end{eqnarray}
they are valid for all $n=0,1,2,3,...$ if one assumes that
$P_{-1}(x)=0$.

Besides the orthogonality condition \eqref{lr0}, we need integrals
involving the derivatives of the Legendre polynomials,
\begin{eqnarray}\label{Ap1p}
\frac{1}{2}\,\int_{-1}^{1}\,P_{n}'(x)\,P_{m}(x)\,dx =
E_{n+m-1}\,\Theta_{n,m+1}\,,
\end{eqnarray}
\begin{eqnarray}\label{Ap2p}
\frac{1}{2}\,\int_{-1}^{1}\,P_{n}''(x)\,P_{m}(x)\,dx =
E_{n+m}\,\Theta_{n,m+2}\ \frac{(n+m+1)(n-m)}{2}\,,
\end{eqnarray}
and in general
\begin{eqnarray}\label{Aplp}
&&\frac{1}{2}\,\int_{-1}^{1}\,P_{n}^{(l)}(x)\,P_m (x)\,dx =
E_{n+m-l}\,\Theta_{n,m+l}\ \times\\
&&\qquad\qquad\frac{2^{-l+1}}{(l-1)!}\,\prod_{s=0}^{l-2}\,
(n+m+l-1-2s)(n-m+l-2-2s)\,,
\end{eqnarray}
where $P^{(l)}(x)$ stands for the $l$-th derivative. Here
$\Theta_{n,m}=P_{n-m}(1)$ is the step function \eqref{step}, and
\begin{eqnarray}
E_{n} = \bigg\{\nfrac{1\quad\text{if}\quad n\ \ \text{is even}}
{0\quad\text{if}\quad n\ \ \text{is odd\ }}\ \ .
\end{eqnarray}
Integrating \eqref{Ap2p} by parts, we have
\begin{eqnarray}\label{Ap1p1}
\frac{1}{2}\,\int_{-1}^{1}\,P_{n}'(x)P_{m}'(x)\,dx = E_{n+m}\
\left[\Theta_{m,n}\,\frac{n(n+1)}{2}+\Theta_{n,m}\,\frac{m(m+1)}{2}
\right]\,.
\end{eqnarray}

\section{Evaluation of $\mathcal{Z}_{k_1 ... k_n}$}

Generally, the coefficient $\mathcal{Z}_{k_1 ... k_n}$ is computed
by taking $n$-th order derivative of \eqref{zint} with respect to
the parameters $s_{k_1}, ..., s_{k_n}$. The result naturally
splits into two pieces
\begin{eqnarray}\label{zksplit}
\mathcal{Z}_{k_1 ... k_n} = \mathcal{Z}_{k_1 ... k_n}^{(*)} +
\mathcal{Z}_{k_1 ... k_n}^{(\text{int})}
\end{eqnarray}
corresponding to two terms \eqref{zint}. Since $x_* = 1+ O(s_k)$,
the term $\mathcal{Z}_{k_1 ... k_n}^{(*)}$ appears only starting
from $n=3$, and in general it has "local" form, i.e. it is built
from the polynomials $Q_0(x), Q_k(x), ...$ and their
$x$-derivatives, taken at $x=1$. On the other hand, the term
$\mathcal{Z}_{k_1 ... k_n}^{(\text{int)}}$ represents explicit
dependence of the integrand in the second term in \eqref{zint} on
the parameters $s_k$; starting from $n=3$, it exists entirely due
to the presence of non-linear terms in the transformation
\eqref{tlambda}. Generally, it has the form
\begin{eqnarray}\label{zknint}
\mathcal{Z}_{k_1 ... k_n}^{(\text{int})} =
\sum_{r=1}^{\left[\frac{n-1}{2}\right]} \
\frac{1}{2}\,\int_{-1}^{1}\, [Q_{l_1 ... l_r}(x)
Q_{l_{r+1}...l_{n}}(x)]\,dx
\end{eqnarray}
where $(l_1, ...,l_n)$ are permutations of the numbers
$(k_1,...,k_n)$, and the symbol $[...]$ signifies the sum of all
{\it distinct} terms generated by the permutations.

Calculation of the "local" term $\mathcal{Z}_{k_1 ... k_n}^{(*)}$
is more tedious since it involves taking derivatives of $x_*$, but
it can be streamlined by reducing it to the residue calculus.
Write
\begin{eqnarray}\label{z*}
\int_{1}^{x_*}\,Q^2(x)\,dx = \oint_{1,\,x_*}\,Q^2 (x)\,\log
\frac{Q\,(x)}{Q_0(x)}\ \,\frac{dx}{2\pi i}\,.
\end{eqnarray}
where the integration contour encircles the points $1$ and $x_*$,
but leaves outside all other roots of $Q(x)$ and $Q_0(x)$. For the
sake of the expansion in $s_k$, we may write $Q(x)$ as
\begin{eqnarray}
Q(x) = Q_0 (x) + {\tilde Q}(x)\,,
\end{eqnarray}
where ${\tilde Q}(x)$ represents all the $s_k$-dependent terms in
the expansion \eqref{qexp}, so that \eqref{z*} takes the form
\begin{eqnarray}\label{zinto}
\frac{1}{2}\,\oint_{1,x_*}\,Q_{0}^2 (x)\,R\left({{\tilde
Q}(x)}/{Q_0(x)}\right)\ \frac{dx}{2\pi i}
\end{eqnarray}
with $R(t)=(1+t)^2\,\log(1+t)$. When $R(t)$ is expanded as the
power series
\begin{eqnarray}
R(t) = t+\frac{3}{2}\,t^2 +\frac{1}{3}\,t^3 -
\frac{1}{12}\,t^4+\frac{1}{30}\,t^5 + ...
\end{eqnarray}
the $l$-th term produces $l$-th order pole at $x=1$ in the
integrand in \eqref{zinto}. Then the local terms are computed by
evaluating the residues.

\bigskip

\textbf{Results}

\smallskip

For $n=3$ one finds
\begin{eqnarray}\label{zkkk*}
\mathcal{Z}_{k_1 k_2 k_3} =
-\frac{Q_{k_1}(1)Q_{k_2}(1)Q_{k_3}(1)}{Q_{0}'(1)} +
\frac{1}{2}\,\int_{-1}^{1}\,[Q_{l_1 l_2}(x)Q_{l_3}(x)]\,dx \,,
\end{eqnarray}
which reduces to \eqref{zkkk} in view of \eqref{qprim}.

For $n=4$ we have
\begin{eqnarray}\label{z4local}
\mathcal{Z}_{k_1 k_2 k_3 k_4}^{(*)}= - \frac{[Q_{l_1} Q_{l_2}
Q_{l_3} Q_{l_4}]\,Q''}{(Q_{0}')^3} + \frac{[Q_{l_1}'Q_{l_2}
Q_{l_3} Q_{l_4}]}{(Q_{0}')^2}-\frac{[Q_{l_1 l_2}Q_{l_3}
Q_{l_4}]}{Q_{0}'}\,,
\end{eqnarray}
where all the $Q$'s are taken at $x=1$, and the brackets $[...]$
have the same meaning as in \eqref{zknint}, for instance the last
term is in fact the sum of six terms.

The integral term in this case has two parts,
\begin{eqnarray}\label{z4int}
\mathcal{Z}_{k_1 k_2 k_3 k_4}^{(\text{int})} =
\frac{1}{2}\,\int_{-1}^{1}\,\left[Q_{l_1 l_2}(x)Q_{l_3
l_4}(x)\right] + \frac{1}{2}\,\int\,\left[Q_{l_1 l_2
l_3}(x)Q_{l_4}(x)\right]\,dx\,.
\end{eqnarray}
The first term here involves only the polynomials $Q_{k k'}(x)$
given by \eqref{qkk}; it is evaluated using \eqref{Ap1p1}, and
when combined with \eqref{z4local}, yields \eqref{zkkkk0}. The
second part in \eqref{z4int} provides \eqref{zkkkkI}.

\section*{Acknowledgments}

We are grateful to P.Di Francesco, V.Kazakov, I.Krichever,
A.Marshakov, G.Moore, A.Polyakov, and M.Staudacher for interest to
this work and discussions. AZ would like to acknowledge warm
hospitality he received at the Bogoliubov Lab at JINR, Dubna, in
the spring of 2008, when this work was done.

AB is supported by the grants RFBR 07-02-00799, and
SS-3472.2008.2. The work of AZ is supported by DOE under grant
DE-FG02-96ER40949.

\end{document}